\documentclass[a4paper]{jpconf}
\usepackage{graphicx}
\usepackage{amssymb,amsmath,caption}
\usepackage{subfig}

\usepackage{graphicx}
\usepackage{tablefootnote}

\usepackage{amsmath}
\usepackage{mathtools}
\usepackage{caption}
\usepackage{footnote}
\usepackage{booktabs}
\usepackage{array} 
\usepackage{url}
\usepackage{color}
\usepackage[section]{placeins}
\usepackage{fancyhdr}

\begin{document}

\title{MeV Neutron Production from Thermal Neutron Capture in $^6$Li Simulated With {G}eant4}

\author{Valentina Santoro$^{1}$, Douglas D. DiJulio$^{1,2}$, Phillip M. Bentley$^{1,3}$}

\address{$^{1}$ European Spallation Source ERIC, SE-221 00 Lund, Sweden} 
\address{$^{2}$ Department of Physics, Lund University, SE-221 00 Lund, Sweden} 
\address{$^{3}$ Department of Physics and Astronomy, Uppsala University, 751 05 Uppsala, Sweden} 

\ead{valentina.santoro@esss.se}

\begin{abstract}
Various Li compounds are commonly used at neutron facilities as
neutron absorbers. These compounds provide one of the highest ratios
of neutron attenuation to $\gamma$-ray production. Unfortunately, the usage 
of these compounds can also give rise to fast neutron emission with energies up to almost 16 MeV.
Historically, some details in this fast neutron production mechanism
can be absent from some modeling packages under some optimization
scenarios. In this work, we tested Geant4 to assess the performance of
this simulation toolkit for the fast neutron generation mechanism. We
compare the results of simulations performed with Geant4 to available
measurements.  The outcome of our study shows that results of the
Geant4 simulations are in good agreement with the available
measurements for $^6$Li fast neutron production, and suitable for
neutron instrument background evaluation at spallation neutron
sources.
\end{abstract}

\section{Introduction}
 
The long (3ms) proton pulse of the European Spallation Source (ESS)
\cite{TDR} gives rise to unique background challenges for the
instrument suite ~\cite{ourpaper1,ourpaper2,ourpaper3}.  In such a
long pulse spallation source, one of the main instrument performance
limits is that defined by the instrument background, feeding into the
Signal to Noise (S/N) ratio.  The instrument requirements for ESS
signal to noise ratio is $>10^{6}$~\cite{bgreqs,bgreqs2} for many
instruments.  Such a signal to noise ratio is achievable, but
absolutely cutting edge at spallation neutron sources, and in some
cases is approaching electronic noise levels.

To achieve the required performance for each instrument, full beam
line simulations must be carried out while taking into account all
beam line components, choppers, shutters, collimator blocks and the
beam dump, with different shielding configurations.\\ \indent Several
different materials can be used in the beam line to shield or to
attenuate the neutron background.  Amongst the different
possibilities, the most used are boron or lithium compounds.  The main
disadvantage of boron compounds is the fact that the attenuation of
the neutron beam is provided through neutron capture in $ ^{10}$B
which releases an alpha particle, a $^{7}$Li ion and a 0.48 MeV
photon. Whilst this gamma photon is considerably easier to manage
compared to those from other absorbers, such as gadolinium or cadmium,
in some cases it is an unwanted background contributor.

Li compounds on the other hand provide very good attenuation for
thermal neutrons with almost no $\gamma$-ray production.
Unfortunately, Li compounds can also produce fast neutrons of energies
up to almost 16 MeV.  The physics process behind this is that, after
being absorbed by $^6$Li, a neutron can induce a triton and an alpha
particle with a cross section of 955 barns at 0.0253 eV
~\cite{radiationbook}. The triton emitted with energy of 2.73 MeV
slows down in the target and interacts with Li and other constituents
of the compound to produce fast neutrons and
$\gamma$-rays. Table~\ref{tab:1} shows a summary of the physics
processes involved.

\indent This fast neutron production mechanism may be missing from
some Monte-Carlo modeling packages often used for shielding design~
\cite{montecarlo}, depending on the optimization being used.  At ESS,
one of the radiation transport codes used for instrument background
and shielding optimisation is Geant4~\cite{geant4package}.  For the
reasons explained above, it is important to assess that this toolkit
is able to describe the fast neutron production process.  To check the
capabilities of Geant4, we simulated the fast neutron production from
Li compounds and compared the results to available measurements from
the following paper \cite{lithiumpaper1}.  In that work, fast neutron
yields from several Li and B compounds were measured and a summary of
the results can be found in Table 2.

\begin {table}[h]
\caption {Fast neutron and $\gamma$ ray producing reactions} \label{tab:1} 
\begin{tabular}{ c c c }
\hline\hline
 Primary Thermal  &  &  \\ 
 neutron induced reactions &  &  \\ 
 Target & Branch +Q(MeV), E$_{ \mathrm max}$(MeV) & Ref \\  
 $^6$Li & $\rightarrow ^4$He + t+ 4.785 $\cdot$ E$_{t}$=2.73  & \cite{tableref1,tableref2}     \\
 Secondary triton  &  &  \\ 
induced reactions at E$_{t} \leq$ 2.8 MeV &  &  \\ 
Target & Branch +Q(MeV),& Ref \\
 $^6$Li & $\rightarrow^9$Be +  $\gamma$ + 17.7   &   \cite{tableref1,tableref2,tableref3,tableref4,tableref5,tableref6}   \\
 & $\rightarrow^8$Be + n+ 16.02   &     \\
  & $\rightarrow$  2 $^4$He + n+ 16.115   &   \\
 & $\rightarrow  ^5$He + $^4$He  + 15.15   &   \\
 \hline
\hline
\end{tabular}
\end {table}

\begin {table}[h!]
\centering
\caption {Fast neutron yields from targets of $^6$Li compounds from thermal neutron capture from the reference paper~\cite{lithiumpaper1} and from Geant4 simulation.} \label{tab:2} 
\begin{tabular}{ c c c  c}
\hline\hline
 Target &  $^6$Li abundance (\%) & measured yield   & simulated  yield \\  
  &  & per $10^6$ neutrons & per $10^6$ neutrons \\  
  &  &  from reference paper   \tablefootnote{The uncertainty in the yields includes the uncertainty due to the incident flux measurement, detector efficiency and any contribution of $\gamma$ background for more details see \cite{lithiumpaper1}. These uncertainty do not include any possible contributions from sample impurity.} &with Geant4  \tablefootnote {This yield number is  obtained using the QGSP\_BERT\_HP  physics list }\\  
  
  $^6$Li & 96 & 84   $\pm$ 26  & 172 $\pm$ 13\\
 LiH& 7.5  & 201 $\pm$ 61  & 82  $\pm$  9\\
  LiF& 7.5 &  178  $\pm$54& 36  $\pm$ 6 \\
 Li$_2$CO$_3$& 7.5   & 107  $\pm$ 33 &  17  $\pm$ 4 \\
 \hline
\hline
\end{tabular}
\footnotetext{} 

\end{table}

\section{Geant4 Simulation toolkit}

Geant4 is a Monte-Carlo toolkit for the simulation of the passage of
particles through matter.  The toolkit is developed by a worldwide
collaboration of physicists, implemented in C++ and has an open source
license.  The code is used across a number of different scientific
fields, including high-energy physics, accelerator physics and medical
and space science.  The physics processes offered cover a
comprehensive range, from meV and extending to TeV energies, and
include electromagnetic, hadronic and optical processes with a large
set of long-lived particles, materials and elements. The Geant4 set of
physics processes to model the behavior of particles are decoupled
from each another (for the most part) and the user selects these
components in custom-designed physics lists that depend on the
particles, the energy range and the process which are to be
simulated.\\
\indent In this work, the Geant4 calculations were carried out using
Geant4 version 10.0 patch 3. The physics lists used for the
calculations performed in this work are the QGSP\_BERT\_HP and
QGSP\_INCLXX\_HP list.  These lists are recommended for shielding
applications~\cite{geant4physics}.  QGS stands for quark-gluon
string, which describes the high-energy interactions of protons,
neutrons, pions, and kaons and nuclei.  The high-energy interaction
creates an exited nucleus, which is passed to the precompound model
(P) which models the nuclear de-excitation.  The label BERT refers to
the Bertini intra-nuclear cascade model \cite{Bertini63,Bertini69},
INCL to the Liege Intranuclear Cascade model ~\cite{incl}, and HP for
the high-precision neutron package \cite{speckmayer}. The HP package
uses evaluated neutron data, named G4NDL4.4, for interactions of
neutrons below 20 MeV. This data largely comes from the ENDF/B-VII.0
libraries. Additionally included in this package is thermal scattering
data for neutrons below 4~eV.

The Geant4 simulations presented in this work were performed with the
ESS Detector Group framework \cite{griff}. The framework is a
Geant4-based Python/C++ simulation environment which contains
developments useful for neutron shielding and detector calculations.

\section{Simulation of fast neutron production from $^6$Li}

\indent To compare the Geant4 results we used a simulation setup that
matches as closely as possible the geometry described in the reference
paper~\cite{lithiumpaper1}, to avoid that our results could be
affected by any geometrical differences.  As a neutron source, we used
$10^6$ mono-energetic neutrons, with a circular beam profile of
diameter 2.5 cm, and with an energy of 0.025 eV.  The $^6$Li target
intercepted the whole of the thermal neutron beam and had a thickness
of 1 mm. This is considered enough for a complete attenuation of the
thermal neutron beam~\cite{lithiumpaper1}.  The lithium sample was
surrounded by a sensitive detector fully covering 4$\pi$ steradians,
to measure the energy of the neutrons coming out from the lithium
sample and to avoid any neutron losses.

\begin{figure}[h]
 \centering
   	\includegraphics[width=12 cm]{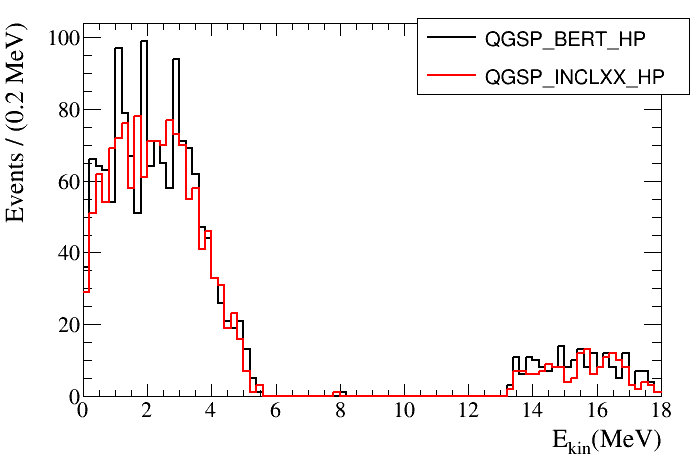}
	\caption[ESS source]{Energy spectrum of the fast neutrons
          produced by thermal neutron capture in $^6$Li for two
          different physics lists.}
	\label{fig:fastneutron}
 \end{figure}

The observed neutron spectrum for the two different physics lists
described above is shown in Fig.~\ref{fig:fastneutron}. This spectrum
has the following interesting features:
 \begin{itemize}
 \item there are several fast neutrons with energies above 14 MeV that
   are consistent with the production mechanism from triton secondary
   interactions as described above.
\item the shape of the fast neutron spectrum is similar to the
  spectrum from available measurements. (See Fig.5. of
  ~\cite{lithiumpaper1})
\item the neutron yields obtained in the simulation (172) for the
  QGSP\_BERT\_HP and (151) for the QGSP\_INCLXX\_HP are on the same
  order of magnitude, ~$10^2$, as the measurements from the reference
  paper (see Table~2).
 \end{itemize}
 
\begin{figure}[h]
 \centering
   	\includegraphics[width=12 cm]{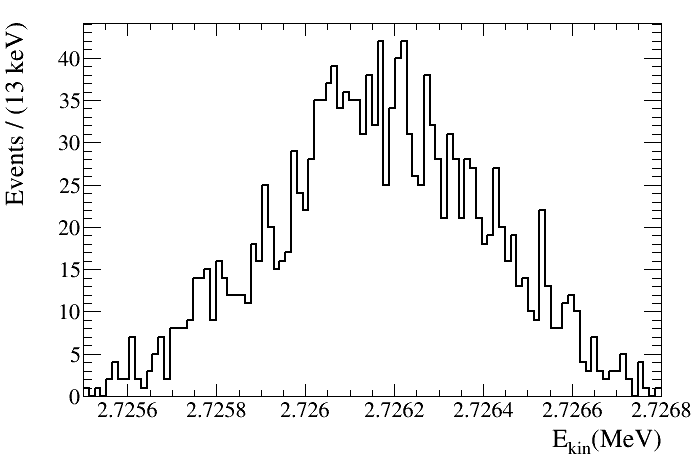}
	\caption[ESS source]{Energy spectrum of tritons produced by
          the reaction $^6$Li(n,t)$^4$He. }
	\label{fig:triton}
 \end{figure}

As an additional check of the physics processes, we also studied the
energy spectrum of the mother particle from which the fast neutrons
are produced.  The study shows that all the fast neutrons were
produced by triton induced reactions. The energy distribution of the
tritons that generated the fast neutrons is shown in
Fig.~\ref{fig:triton}.  The energy spectrum is consistent with tritons
being produced from interactions of a thermal neutrons with $^6$Li and
the spectrum is almost monoenergetic due to conservation of energy and
momentum.  In summary, all the features illustrated above show that
the primary and secondary reactions are described reasonably well in
the Geant4 simulation package.\\

 \section{Lithium compound simulations }
 
\begin{figure}[h]
 \centering
   	\includegraphics[width=12 cm]{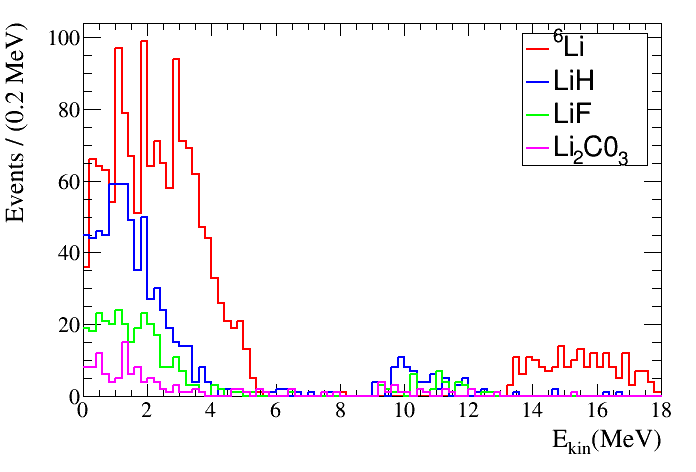}
	\caption[ESS source]{Energy spectrum of the fast neutrons
          produced by thermal neutron capture in different Lithium
          compounds}
	\label{fig:lithiumcompounds}
 \end{figure}

For completeness, we also tried to simulate several different Li
compounds to compare with the expected yield from Table~2.  The
resulting spectra are shown in Fig.~\ref{fig:lithiumcompounds}. The
shape of the spectra for LiH, LiF and Li$_{2}$CO$_{3}$ seems
consistent with what is shown in the reference paper, but the yields
are somewhat lower than expected (see Table 2). The differences
between the measured and simulated spectra may be due to a number of
factors.  These include the lack of precise thermal neutron scattering
data for some of the materials of interest; geometric and material
composition differences that are not fully documented; and the
reliance on nuclear models for the charged particle interactions. This
last point may be addressed with an up and coming package, titled
G4ParticleHP~\cite{scatteringthermal}, for Geant4. This package will
introduce evaluated data for simulating the interactions of charged
particles in Geant4. In order to further
understand the origins of the differences between the measured and
calculated spectra, a detailed investigation of the multi-step process
would be necessary for each material.

\section{Conclusions}

We have presented simulations of the fast neutron production channel
of lithium absorption of thermal neutrons using Geant4.  The
simulations compare favourably with available measurements for $^6$Li
fast neutron production in the literature.  As such, we conclude that
sufficient sensitivity and reliability exists in Geant4 for full beam
line models of different instruments that are planned at ESS, where
$^6$Li absorbers are the primary means of achieving efficient neutron
shielding.

\section*{References}
\bibliographystyle{iopart-num}

\end{document}